# INSTITUTIONAL SIMILARITY DRIVES CULTURAL SIMILARITY

# AMONG ONLINE COMMUNITIES


Qiankun Zhong[1], Seth Frey[1]

[1]Department of Communication, UC Davis

Direct correspondence to Qiankun Zhong, qkzhong@ucdavis.edu




**Institutional Similarity Drives Cultural Similarity among Online Communities**

**Abstract**

Understanding online communities requires an appreciation of both structure and culture. But basic questions remain difficult to pose.  How do these facets interact and drive each other? Using data on the membership and governance styles of 5,000 small-scale online communities, we construct empirical measures for cross-server similarities in institutional structure and culture to explore the influence of institutional environment on their culture, and the influence of culture on their institutional environment. To establish the influence of culture and institutions on each other, we construct networks of communities, linking those that are more similar either in their members or governance. We then use network analysis to assess the causal relationships between shared culture and institutions. Our result shows that while effects in both directions are evident, there is a much stronger role for institutions on culture than culture on institutions. These processes are evident within administrative and informational type rules.

**Institutional Similarity Drives Cultural Similarity among Online Communities**

It is widely appreciated that the culture and institutions of a social system are deeply related. The connection is frequently discussed by communication scholars (Cheney, 1983; Stohl & Cheney, 2001), political scientists (e.g., Cook, 1998), sociologists (e.g., Zucker, 1977), anthropologists and economists (e.g., Henrich et al., 2001). And yet, a general understanding of their relationship has remained elusive. The challenge of generalizability in the cultural studies approach (Barnett, 1997) and the differing theoretical focus of other co-evolutionary organizational research (Monge, Peter & Contractor, 2003) result in a gap in understanding the mutual influence of culture and governance institutions in communication. Online communities provide a unique opportunity for communication researchers to observe their interactions and illuminate the processes of this co-evolution by facilitating comparison among a large number of separate organizations across time (Hill & Shaw, 2017).

An organization can be seen as a system composed of a set of interdependent components, including individuals, groups, or machines, structured by constraints and goals (Goldhaber & Barnett, 1988). Organizational communication focuses on many types of these constraints, such as information processing capacity (Tushman, 1977),  symbiosis and competition with other organizations (Barnett, 1990; Brittain, 1994 )  and environmental circumstances (Thompson, 1967); and many of the effect of these constraints on an organization's members, such as the degree of access to shared information and differentiation of individual positions (Burns & Stalker, 1961); and even how these drive each other, such as environmental processes (Baum & Korn, 1996) and demographic processes (Freeman & Hannah, 1983) that foster the coevolution between the population and organization.



But an "emergent" account of organizational outcomes in terms of individual behavior requires a deeper understanding of "institutional" dimensions of organizational processes. The type of person who joins an organization must have a large effect on the rules, norms and enculturation processes it develops. These rules, norms and enculturation processes must have a major influence on who chooses to join ("select into") the organization.

This view of organizations is in line with Simon's administrative behavior model and North's institutional theory. Simon (1976), while less interested in culture *per se*, claims that rational behavior in organizations can be attributed to goal specificity and formalization. Work in Simon's tradition has, for example, examined how the rules of an organization evolve over time in conjunction with the evolving needs of its membership (March, Schulz & Zhou, 2000). North (1991) considers institutions as a set of formal and informal constraints that structure individual behavior. We draw heavily on the work from both: linking boundedly rational agents (administrative) to the type and role of behavioral constraints (institutional) in determining how a population evolves.

But quantitatively establishing and characterizing these mutual linkages, down to their relative strengths, is a major research challenge.  At minimum, it requires rich data on the structure of a large numbers of organizations, and their rules systems, that are linked to each other by a large and diverse population of potential members. We study a large "population of populations": a system of interlinked online communities that cluster in terms of drifting cliques of shared users.  Over several months, we observe both the patterns of user traffic between them and the evolving governance systems of each community.  This permits us to observe the coevolution of social niches (sharing what we represent here as a cultural identity) and structure (which we understand here in terms of governance style).



Previous research in this domain (Frey & Summer, 2019) applied the framework of "institutional analysis and development" (Ostrom, 2005) to this setting and found a positive correlation with communities between governance complexity and community success. However, this research disregarded the evolution of communities over time, as well as their interactions with each other. We focus here on the reciprocal questions of whether servers with the same users converge on the same governance styles, and whether servers with similar governance styles attract similar users.

## Literature Review

### Institutions and Culture

**Definitions of institutions and culture** There have been many attempts to define institutions and culture, many of which articulate a tacit relation between the two. North (1990) defines institutions as "the humanly devised constraints that structure human interactions. They are made up of formal constraints (i.e., rules, laws, constitutions), informal constraints (i.e., norms of behavior, conventions, self-imposed codes of conduct), and their enforcement characteristics". In North's theory, formal rules are created by the policy, whereas informal norms are "part of the heritage that we call culture." By contrast, Greif (2006) defines an institution as "a system of social factors that conjointly generates a regularity of behavior." Greif refers to social factors as "man-made, nonphysical factors that are exogenous to each person they influence," including "rules, beliefs, norms, and organizations." Both allow that culture and institutions are endogenous social factors. And both suffer from a failure to properly disentangle these constructs: both "institutions" overlap too much with culture. For example, they each define both institutions and culture in terms of "norms" and "conventions" (Alesina & Giuliano, 2015). Thus, when measuring cultural and institutional structure, cultural economists usually



distinguish institutions and culture in terms of the formal and informal, or written and unwritten portions of a society's system of order. Political scientists take a more micro view, considering political culture as the "theory" of an individual's theory of what his/her fellow knows, believes, and means (Keesing, 1974). In this sense, culture is interpersonal and consists of a set of theories. In line with this approach, Geertz considers culture to be a set of control mechanisms rather than a complex entity of concrete behavior patterns. Sociologists note that different thinkers pay attention to distinct aspects of culture and conclude that a set of overlapping factors constitutes culture. Swidler (1986) and Tilly (1993) proposed a framework defining culture as a repertoire that includes symbols of meanings and practices selectively used by group members to construct collective action strategies and collective identities. This framework provides an approach for us to distinguish culture (toolsets for individual behavior) from institutions (constraints on individual behavior). The cultural repertoire framework also allows us to connect individual behavior with group strategy and group identity. Therefore, we here rely on Swidler's and Tilly's framework.

**Institutional influence on culture** Having considered how a social system's institutions and culture are distinct, it becomes possible to theorize how they affect each other. Institutions forcefully shape how people act and interact with one another and interpret those actions. They write people into specific identities and social categories that can persist for a lifetime. The influence on people's experience shapes their perception of who they are, how they should behave, and their relationships with others within their communities. In this sense, institutions are powerful forces that shape culture. Research on the institution of the Burning Man festival (Chen, 2005) provides an example of how core institutions of Burning Man have a lasting effect on Burners' social activities. Past participants were observed to start supporting public art at



regional events in their local communities, evidence of a cultural diffusion process that was triggered by institutional factors. Bowles (1998) surveys a broad array of evidence on the cultural effects of market institutions and market exchange. Tabellini (2010) provides empirical evidence of formal institutions affecting the evolution of cultural traits by identifying that European regions with less well-developed institutions in the past have less trust and less respect for others and believe less in the value of individual effort today.

**Cultural influence on institutions** The influence of culture on institutions is usually explained by the influence of certain cultural factors on individual behaviors or perceptions, which foster the historical development of institutions (e.g., Licht, Goldschmidt and Schwartz, 2005) or affect the institutional results in certain cultural context (e.g., Guiso, Sapienza, and Zingales, 2016). Most empirical research on this matter focuses on the effects of cultural factors including cultural-psychological difference (e.g., individualism versus collectivism), general morality (e.g., Tabellini, 2008), family ties, and perceptions of equality and efficiency (e.g., Alesina, Galeser, and Sacerdote, 2001).

## Online Communities for Relating Institutions and Culture

Alesina and Guiliano (2015) emphasize the closeness of the link between culture and institutions by arguing that causal linkages between them must flow in both directions in a feedback relationship that drives larger system-level dynamics.

However, it is often prohibitively difficult to rigorously analyze a social system's culture and formal institutions together. First of all, the definition of culture and institution varies in different fields (Alesina & Guiliano, 2015), which also leaves competing measurements of both. Second, it is hard to isolate the cultural influence on an institution or the institutional influence on cultural development, given that they are both endogenous social systems. This also causes



methodological difficulties in examining their relationships. Field experiments can establish causal relationships between culture and formal institutions (e.g., Bapna & Umyarov, 2015; Chen, Harper, Konstan, & Li, 2010). However, large-scale field experiments are usually expensive and difficult to coordinate. Moreover, when scholars do conduct field experiments, they can be limited by small sample sizes (Harris, 2018).  Observational studies can also address the relationships between culture and institutions by comparing two communities, but it is rare and remarkable when we encounter two communities with similar enough features to infer causality from their differences (Castronova, 2006).

Social systems that exist online make this type of research much more feasible. Among other things, they cost significantly less than large-scale observational studies—making it easier for interdisciplinary scholars to engage with large populations (Hill & Shaw, 2017). Online social systems that facilitate semi-independent online communities, including *Minecraft* and *Reddit,* are positioned to facilitate comparisons between communities because each sub-community has (1) identical initial conditions, (2) public history and documentation, and (3) sovereignty in the founding and developing of institutions and culture. These systems encompass communities that are similar enough to compare with one another, allowing researchers the opportunity to study large-scale collective behavior with low cost and high efficiency. In this study, we track several thousand *Minecraft* servers over five months with a focus on proxies for governance and culture to analyze how these dimensions of a community interact.

**The Case of *Minecraft***

Interestingly, the video game *Minecraft* provides a useful context for discussing the relationship between subcommunities and their rules. *Minecraft* is a massive multiplayer online game that allows for various autonomous user activities, including building with blocks,



exploring a virtual world, gathering resources, exchanging goods, and engaging in game combat. In *Minecraft*, users can establish and manage their own servers for playing the game. These servers function both as spaces other users can explore and communities that users can engage with (Frey & Summer, 2019). Someone who sets up a server, the system administrator, takes the responsibility of governing it. To achieve success at building a community of players around their server, administrators have to recruit and retain repeat visitors who can and do migrate between servers at a low cost. Administrators also face constraints in physical resources (e.g., RAM, CPU, bandwidth, monthly server fees) and virtual resources (e.g., software-based currency, reputation systems), all of which must be carefully managed to provide the membership with a quality game experience.

**Institutions and *Minecraft* Plugins** In the *Minecraft* ecosystem, administrators rely on custom software-based governance institutions to manage limited resources, solve collective action problems, and eventually maintain a corps of quality community members. These "plugins" are modular programs that administrators can install on their servers to automatically implement rules and other political-economic constructs. Plugins can allow for certain behaviors or activities, or improve the experience of them. For example, "parties" is a plugin that allows administrators to hold parties in their community. Others prohibit abusable behaviors or make it easier to administer punishments for rule violations (e.g., the "AntiCheat" plugin prohibits cheating behavior in the game, while "Combatlog" punishes players who unfairly evade the consequences of starting a losing battle). By "mixing and matching" plugins and fine-tuning their settings, server administrators craft highly customized formal institutions and implement a social structure that can solve problems and achieve governing goals. In *Minecraft*'s setting, players can switch between servers at a low cost, which leads to fierce competition between servers. To



recruit and retain members, administrators have to implement plugins that benefit players' experience.

So far, the *Minecraft* community has developed almost 20,000 plugins. To assist administrators in selecting and using the plugins effectively, the software requires plugin developers to assign each plugin to at least one pre-specified category. As of 2016, when our data collection ended, the *Minecraft* developer community listed 16 types of plugins administrators could use to implement rules[1]. Among those, Frey and Summer (2019) identified four types that directly related to governance: top-down administration, communication, economy, and information.

Plugins in the administration category allow administrators to execute additional control over server states and player behavior toward preventing or remediating problem behaviors among the games anonymous and young users.  Top-down administrative control mechanisms may strengthen founders' role identities and commitments to the community, ultimately leading to additional governance efforts (Butler, Sproull, Kiesler, & Kraut, 2002). Specific administrative control tools (e.g., responsibility systems) have been shown to increase the trust between a community's members (Ba, 2001). However, some argue that centralized control inhibits knowledge sharing efficacy (Godara, Isenhour, & Kavana, 2009) and collective judgment accuracy (Becker, Brackbill, & Centola, 2017).

Plugins in the communication category facilitate interpersonal communication by providing additional or higher bandwidth channels for peer-to-peer communication. Institutional communication mechanisms have been positively correlated with engagement (Brodie, Juric, &

---

[1] The plugin categories are Admin Tools, Anti-griefing Tools, Chat Related, Developer Tools, Economy, Fixes, Fun, General, Informational, Mechanics, Role Playing, Teleportation, Website Administration, World Editing and Management, World Generators, and Miscellaneous. See
https://www.curseforge.com/minecraft/bukkit-plugins/world-editing-and-management.



Hollebeek, 2013), customer bonding (Szmigin, Canning, & Reppel, 2005) and the intent to participate in the online community (Wise, Hamman, & Thorson, 2006).

Economy plugins protect private property rights and facilitate resource exchange. The economic institutions in large games are unique because they usually have built-in virtual currency systems. However, these same types of resources and incentive systems exist in almost all online communities in the form of knowledge and social capital. Although researchers generally recognize knowledge and social capital as key resources in online communities (Sharratt & Usoro, 2003; Chiu, Hsu, & Wang, 2006), minimal scholarly work investigates resource governance and its influence on communities. Harris (2008) uses Ostrom's (2005) framework to propose institutional solutions to free-riders in a peer-to-peer network based on peer provision of scarce goods.

Finally, informational plugins provide more channels for broadcasting messages and regulations to the community. In contrast to the peer-to-peer communication facilitated by communicative plugins, informational plugins promote top-down communication. Kraut and Renick (2012) found that it is crucial for communities to clearly state rules, expectations, and affordances, particularly for the purpose of enculturating new members.

**Culture and *Minecraft* Plugins** Drawing from one element of the sociological conceptualization of culture, we understand the culture in Minecraft as a set of cultural repertoires based on shared practices and experiences, and isolate it behaviorally in terms of the server communities that players self-select into: a major part of the ecology of the game is that players can choose what community they join. A cultural repertoire is acquired through players' experiences in different servers and interactions with other players. Communities can identify with "tags," markers that a community has self-identified into one of several game subcultures.



For example, tags named "Pokemon," "Hunger Games," or "PvE" indicate, respectively, affiliation with *Pokemon* or *Hunger Games* roleplay, or a focus on "player versus environment" play (rather than "player versus player"). Tags, as a facet of culture, sustain the stability and functioning of communities by broadcasting shared meanings and commitments that bind individuals into groups (Durkheim, 1915). The communities that a user selects into thus give a sense of the range of identities that the user holds. And when many users share overlap in their range of identities, evidenced by their tendency to traffic in the same subset of servers, we infer that they share this sense of culture, operationalized here both as overlap in the set of subcultures, and frequent interaction over a set of similar communities. To be clear, we are not attempting here to define the cultural identity of users in terms of the communities they visit, nor the subculture of a community by the users who visit it: we are attempting to define a set of communities as culturally similar by the existence of a large and consistent group of users who tend to travel between them.[2]

In *Minecraft*, both culture and institutions are less persistent and change much faster than in the real world, due to servers' short lifespans and players' fast-pace gaming activity, which includes interaction, creation, and destruction. Therefore, *Minecraft* makes it easier to witness the co-evolution of culture and institutions over timescales as small as weeks and months.

**Hypotheses**

---

[2] Of course, it is possible that a member regularly travels between two communities that have different cultures, understanding the distinctive social norms of both, and adjusting accordingly when shifting between the two. Evidence on online identity demonstrates substantial fluidity across media and communities (e.g., Leavitt, 2016; Leavitt, Clark & Wixon, 2016).  But even allowing this, the fact that a coherent and sizeable group of people is showing the same pattern of fluidity between the same two distinct cultures, in preference to any other pair of communities they could be traveling between, is difficult to explain without some concession that those communities share something cultural in common.



Based on our definition of *Minecraft* culture — repertoires of shared meaning that provide references for preference and behaviors — it is reasonable to measure the cultural similarity between servers by dual membership; servers with a high share of dual membership will share a similar cultural repertoire either because, having advertised on similar markers of identity, they attract the same types of users, or because of the knowledge and practices that their dual-membership users contribute to both group repertoires. Accordingly, servers with a low share of dual membership are less likely to share a similar group repertoire because their members are less likely to have similar identities or experiences in *Minecraft*.

Given this close connection between shared membership and server culture similarities, two claims become apparent. On one hand, member culture can affect institutional development through user preferences. Server success is almost solely dependent on recruiting and retaining repeat visitors, so users indeed have the bargaining power to negotiate with administrators about institutional decisions. Through this kind of mechanism, users' preference can possibly influence institutional development. On the other hand, *Minecraft* institutions can affect a member culture through their influence and limitations on member interactions. *Minecraft* governance institutions have a forceful effect on how visitors behave and interact with each other in the server, which also influences visitors' self-image, values, and preferences. Accordingly, people with the same experiences, shared values, and norms will be more likely to be bound together. In other words, regular visitors of one server may be more likely to visit servers with similar institutions. Thus our two motivating hypotheses,

*H1: Shared membership between Minecraft servers causes them to become more institutionally similar.*

*H2: Institutional similarities cause shared membership between Minecraft servers.*



## Data

### Server data

We analyze longitudinal user visit data from 370,000 *Minecraft* servers contacted through API queries bihourly between Nov 2014 and Nov 2016. The scraper accessed each server for their visitors' anonymous user ID and visit times, plugins installed, and other server features. An important feature of the *Minecraft* ecosystem is that user IDs persist across servers, making it possible to observe a user's trajectory across many otherwise independent communities. Following Frey & Sumner (2019), we first filtered out servers that were disconnected for the duration of data collection (~220,000), those that did not survive for at least a month (~70,000), and those that did not report full governance information (~75,000). We then further refined the resulting 5,215 servers in order to create the minimum conditions for the viability of our analysis (we address the potential effects of bias due to non-random deletion of data in the Limitations). To capture network formation and evolution continuously, we selected servers that were live for over 16 weeks in a 5-month timespan. Our analysis required us to aggregate over weeks to make the month the basic temporal unit of analysis. As the median "lifespan" of a server is 9 weeks, and administrators pay for server space monthly, the timescale of a month gives a suitable level of granularity. To match the timing of visit data on each server with the timing of rule changes, we selected servers that were operating between week 5 and week 22 in 2016 and had a final corpus of 1097 online server communities.

Among these servers, we identified 2,791 unique plugins in use in the four governance plugin categories: 1,310 administration plugins, 520 communication plugins, 369 economy plugins, and 735 information plugins.

## Method



To quantitatively pose our questions about mutual influence, which are dynamic and involve several types of relationships between many communities, we resorted to a recent network analysis approach, the dynamical multiplex spillover method (Vijayaraghavan, Noël, Maoz, & D'Souza, 2015). We use dynamical multiplex spillover to represent governance similarity relationships and membership similarity relationships as different networks over the same nodes ("multiplex"), and quantify the effects of link additions and deletions in one network on the corresponding links in the other ("dynamical"), relative to a theoretical null model that estimates baseline probabilities of links appearing and disappearing. Multiplex networks, also known as multilayer networks, provide a way of representing a system when nodes can be linked by links with different meanings.  For example, Szell, Lambiotte, and Thurner (2010) take a multiplex approach to link entities in terms of ties denoting several different kinds of relations, including friendship, enmity, and economic partnership.

The dynamical multiplex spillover is well suited to exploring the causal pathways between (1) cultural similarities between servers, measured by *shared membership*, and (2) institutional similarities between servers, measured by *shared server rules,* servers' similarities in their extent of use of each institutional plugin type. Under this method, if links representing, for example, governance similarity show significant changes after a change in their membership links, that would constitute evidence in support of the hypothesis that cultural factors drive institutional factors.  Of course, dynamical multiplex spillover models do not offer experiment-quality causal inference. But their use of change over time in combination with a well-formulated null model of change puts them above, for example, a design based on a simple pre/post comparison.



Using longitudinal data on shared membership traffic and servers' rules, we first constructed a network of five layers (*Figure 1*), combining *shared membership network* dynamics with four kinds of *server rule network*, corresponding to each of the four types of governance institution: administrator-focused mechanisms, chat mechanisms, information distribution mechanisms, and economic mechanisms. In our network, nodes signify servers, whereas links can have different meanings depending on what layer they are in. By analyzing the network processes both within and across layers in the multiplex, it becomes possible to use social network statistics to represent the idea that cultural and institutional processes interact.

We construct the shared membership traffic measure by calculating the number of users who visit two or more servers within the same month. We then dichotomize the shared membership traffic by a median split, satisfying a constraint imposed by our statistical approach, which cannot leverage continuous link weights. The users with shared server membership bring to both servers their experiences and practice, which weighs in the creation and development group cultural repertoires in both servers. Therefore, servers with a high share of mutual visitors should be more similar in culture, compared to those with a low share of mutual visitors. At the same time, we use server rules to measure shared server rules. A server's relative preferences for rules in different categories provide a proxy for its style of governance. The dynamics of rule establishment, including increasing or decreasing the number of rules of each type, proxies institutional development within each server. For each server, we create dummy variables for each of the four types of rules to characterize the servers as either high or low in each of the four types of rules. We determine "high" and "low" on the basis of a median split.

In the community network, the presence of a link indicates that two servers have a high number of shared members. In the rule networks, the presence of a link between two servers



indicates that the two servers have implemented a similar amount of rules of that type (i.e., similar numbers of administrative, informative, communicative, or economic plugins; see *Figure 1* below).

If culture and institution have no effect on each other, the dynamics of link appearance and disappearance in the five-layer multiplex network should be indistinguishable from the link dynamics of the five networks considered independently. On the other hand, if rules have an effect on culture, or vice-versa, we should observe that when links in the rule (or shared membership) network change, links in the shared membership (or rule) network change at a different rate than that expected from two statistically independent networks. The method draws these null predictions from the statistics of Markov chains, a well-understood theoretical model of how systems change over time.

[Figure 1 about here]

Our approach allows quite fine-grained investigations, permitting us to ask not only what kinds of changes in one elicit changes in the other, but also the strength and direction of these interrelations, whether an increase, decrease, or persistence of links at one layer is associated with an increase, decrease, or persistence of links at another layer. This approach also required differentiating between the incremental, sequential "slow-timescale" transitions and more sudden "fast-timescale" transitions. In the type of slow time-scale transition at the root of our inquiry, a pair of nodes linked at only one layer might transition to being linked at two. By contrast, in fast-timescale transition, a pair of nodes linked at neither the membership or rule layers are in the next time-step linked on both. When a dyad changes slowly from having high similarity in one layer to having high similarity in both that and the community layer, we can interpret that within one of our directional hypotheses. But when a dyad changes from being low similarity on both



layers to high on both, any direction of influence between the layers is impossible to discern. Measuring both slow-timescale and fast-timescale transition gives a full picture of the co-evolution dynamics in *Minecraft*. Of course, this same flexibility meant an explosion in the number of statistical comparisons we could perform (4 types of change in shared community link (*continued presence, continued absence, appearance, disappearance,*) × 4 types of rule link change × 4 types of rule = 64 potential comparisons), a problem we use theoretical constraints and specific hypotheses to tame  (*Figure 2*).

## Result

To test our hypotheses, we identified the causal effects in multiplex networks by comparing observed patterns of multi-layer link dynamics to those predicted by the null model (Vijayaraghavan, Noël, Maoz, & D'Souza, 2015).

At each time step, for each pair of nodes, for each type of link, a link may appear, disappear, persist, or continue to not exist, for each direction of influence, signifying in each case the coevolving similarity or dissimilarity of two communities. Looking over the institutional and cultural networks simultaneously, there are 16 possible transitions (*Supporting Figure 1)*, of which we take an interest in just a few (*Figure 2*). To capture institutional links each appearing, disappearing, persisting, or remaining absent, we describe dyads in 4 possible states: being high similarity on both institution and culture, being low on both, or being high on one and low on the other.

The diagram highlights some transitions that are theoretically important and others that are not, but that must be estimated as part of the model. It also illustrates the "adirectional" fast-timescale changes between network layers that are impossible to view as being driven by one or the other layer (F1.1 and F1.2 in *Figure 2*). We begin by investigating the fast-timescale changes



to motivate the more directional and theoretically pertinent "slow-timescale" changes of central interest to this work: the spillovers from culture to governance and governance to culture; the probability that two servers will gain in institutional similarity given that they share cultural similarity (Hypothesis 1, captured by S2.1 having a positive value in *Figure 2*), and the probability that they will gain in cultural similarity given they already share institutional similarity (Hypothesis 2, captured by S1.1 having a positive value in *Figure 2*).

To calculate the effect size of cross-layer spillover effects, we calculate the difference between observed and null probabilities of transitions in Figure 2, which, on a technical level, can be interpreted as a Markov chain whose transition probabilities (arrows) between states (boxes) are both empirically and theoretically calculable.  We interpret a positive (or negative) spillover as occurring between two layers when the 99% theoretical confidence interval around the difference between observed and null probabilities excludes a difference of zero.

[Figure 2 about here]

Looking at the fast-timescale co-transitions, institutional and cultural links appearing or disappearing simultaneously (F1.1), there is a general trend from low to high in both shared membership and rule similarity for two types of rules: the difference between the observed and null probability of transition F1.1 is greater than zero with at least 99% confidence for both administrative rules ($F1.1_{diff.}$ = 0.0007[0.0004, 0.0010]; square brackets indicate 99% confidence interval around all statistics; see *Figure 3*) and informative rules ($F1.1_{diff.}$ = 0.0008 [0.0005, 0.0011]). The other two rules types were insignificant: the observed/null differences were not distinguishable from zero for economy rules ($F1.1_{diff.}$ = 0.00018 [–0.0003,0.00066]) and communication rules ($F1.1_{diff.}$ = 0.0000003 [–0.0000005,0.0000011]). In other words, dyads that gain cultural or institutional links tend to simultaneously gain the other types of links at a higher



than expected rate for some types of rule. These results, while encouraging for theories that posit

relationships between culture and institutions, does not give any hints as to directionality.

[Figure 3 about here]

The slower timescale processes evident in other transitions (S1.1, S1.2, S2.1 and S2.2 in and then became similar on the other, are more amenable to directional or causal interpretation.

*Figure 1*), in which two servers were already similar on one dimension (culture or institution)

For example, S1.1 represents two servers with high rule similarity transitioning from low shared

membership at one timepoint to high shared membership at the next timepoint. This transition is

significantly above its expected value (*S1.1$_{diff.}$ = 0.146 [0.036, 0.256]; See Figure 4*), whereas

the spillover effects in S2.1, indicating the reciprocal effect of shared membership similarity on

rule similarity, are not significant (*S2.1$_{diff.}$ = 0.068 [–0.052, 0.188])*, indicating that cultural

similarities are not strong driving factors of institutional similarities.  As a corroborative test of

our directional hypotheses, we also check the probability that two servers similar on both

dimensions will stop being similar on one dimension (S1.2 and S2.2 in *Figure 2*). Our

hypotheses predict a decrease: that these transitions will be observed significantly *less* often than

would be expected by the null model. We find servers that are similar in one dimension are less

likely to stop being similar in another dimension ( *S1.2$_{diff.}$ = –0.147 [–0.351, 0.056]; S2.2$_{diff.}$ = =

–0.151 [–0.275, –0.028]; Supporting Figure 3*). This indirect effect, although not strong

evidence, is ultimately consistent with a positive feedback effect.

[Figure 4 about here]

Repeating the above analysis on the other three rules types, we find the same asymmetric

pattern of two servers with high rule similarity transitioning from low shared membership to high

shared membership (*S1.1$_{diff.}$ = 0.124[0.029, 0.219]*) and marginal reciprocal effects of two



servers with high shared membership transitioning from low rule similarity to high rule similarity ($S2.1_{diff.} = -0.306[0.0479, 0.1091]$) for information-type rules. For communication- and economy-type rules, we find no effect in either direction (See Figure 5).

[Figure 5 about here]

To summarize, these results suggest that online community's cultural and institutional dimensions are interlinked in several senses: a pair of communities that become more similar on one are significantly more likely to simultaneously become more similar on the other, and servers that are already similar in terms of governance are significantly more likely to become more similar on culture as well. All of these results hold for two rule types only: administrative and informational, with no deviations from null for any of the transitions linking economic and communication rules to the community network. Although there are many interpretations of these results, H1 is the hypothesis that is most consistent with what we observe: that institutional similarities increase patterns of shared culture among *Minecraft* servers. That said, our hypotheses are not mutually exclusive, and we also find some evidence consistent with H2, namely that communities that are already similar on both governance and culture are slightly less likely than expected to become dissimilar on governance in the next time interval.

## Discussion

### Findings and Contribution

In this study, we support theories positing a positive feedback loop between culture and institution, finding, however, that institutions in *Minecraft* communities have stronger effects on culture than does culture on institutions. Specifically, we show that communities that govern themselves with similar types of rules are more likely than expected to subsequently attract



similar users. Our approach to this challenge not only reveals interactions between culture and institutions but also shows the dynamic processes occurring over different timescales.

In this research, we study culture as an endogenous factor, which allows us to theorize and test how it co-evolves with institutions and to measure it by user visits. Therefore, we have to take a functionalist view to analyze the impact of culture. That is, we view society (in this context, online *Minecraft* communities) as a complex system of individuals and organizations that work together to promote the stability and functioning of the whole. Functionalism was criticized for focusing only on social order and being insufficient for accounting for social change. In our case, there seem to be some theoretical conflicts between functionalism and cultural evolution, which sees culture as a dynamic process that adapts to the environment. It is true that although functionalism leaves space for social change (Durkheim & Halls, 1986), it doesn't explain the changing process and mechanism very well. In this study, we use cultural evolution to fill in the gap in functionalism and understand the social structure as an evolutionary process, which accounts for both evolution and stability. Another criticism of functionalism is that it downplays the individual's agency in society. We reveal similar limitations in our measurement and analysis that we overlook individual motivations other than culture that drive users to visit different servers at the same time. Indeed, research in complex systems uses statistical mechanics to emphasize the macro-scale regularities that emerge reliably and independently of the individual-level dynamics (e.g., Castellano, Fortunato & Loreto, 2009; Fortunato, Macy & Redner, 2013; Kollman & Page, 2006).

Here we measure culture by shared user visits regardless of all other possible motivations because our focus is on cultural similarities rather than the actual content of culture per se. For example, interpersonal connections and popularity may lead users to servers where they know



those members already. However, this drive does not change the fact that the same population (shared members) still share the same experience and practice in *Minecraft*, and those shared experiences still construct similar group culture.

Another interesting finding of this research is that we only observe spillover effects among administrative and informational rules. The reason might be that the two types of plugins are more relevant to new members, who are also the population that are more likely to migrate between servers, whereas communicative and transactional rules are more likely to be useful for those who are already familiar with the server and more stable as members of the communities.

## Limitations

Regarding the measurement and data, our analysis used five months of server and user dynamics, which might seem not long enough to capture the transition and learning in *Minecraft*. Within the five months, servers might be at different stages of their life. Some servers might be at a more stable state than others, whereas some might be expanding at the time and eager to get experiences from successful neighboring servers who share members with them. Indeed, it is possible that the five months only covers certain parts of servers' life. However, the median life of *Minecraft* servers is nine weeks. Given that *Minecraft* servers evolve quickly, our assumption still holds that our data capture transitions and learning over the course of a server's life. It is also possible that the effects we observed among particular categories are due to exogenous temporal factors. Mostly, the null model we constructed already takes into account exogenous factors and produces direct correlations between different layers of networks.

In addition, although this paper provided causal pathways to the relationships between institutions and culture, it still cannot make true causal inference because it is not an experiment with random assignment. We cannot control for time-varying network-specific variables that



might also influence network processes. For example, non-stationarity might affect network dynamics, which would undermine the result. Other hidden instrumental variables that correlate with changes in one layer but not with another may also alter our results.

Constraints of the method made it difficult to include in our model the effects of overall community size on shared membership. It is possible that communities with a larger size are more likely to have shared membership with others over time, which may confound the network processes.

Another constraint of our method was its complexity, which, in addition to constraining its general usability by others, resulted in many more parameter fits and statistical comparisons than we anticipated, or that we could theorize over. As a result, we engaged in *post hoc* theorizing that ultimately influenced our formulation of the hypotheses, in what may be a dangerously circular manner, as well as which of the many results to highlight. Although a mix of exploration and hypothesis testing is to be expected to some extent in non-experimental research applying new methods in new problem domains, it should at least be pointed out. Future work should attempt to reproduce our findings in the course of extending them, rather than merely taking them for granted. We report full results over all parameters in the Supporting Figures.

## Implications and Future Studies

We examined institutional development in *Minecraft* and its interaction with culture, which provides a framework for future institutional analysis and computational studies of cultural dynamics. The different effects of rule types provide an opportunity for future research to identify the effectiveness of specific rules. In this research, we found significant effects among administrative and informational rules, but no effects from communicative or economic rules.



This result motivates valuable research questions, including how do different types of rules work? Who do they affect the most? Moreover, why do players hold preferences toward different types of rules? The focus on different rules marks the institutional difference between *Minecraft* servers, which provides us several dimensions to evaluate and compare the organizational institutions.

Another direction that might inspire future research is the cultural aspects of online communities. As mentioned earlier, we study culture as an endogenous factor. Future studies can take another approach by studying the exogenous cultural effects on institutional development. Cultural indicators such as language and server IP address might leave us hints on how exogenous cultural factors influence the collective behavior and institutional development in *Minecraft* communities.

Last, we highlight a fundamental advance that made it possible for us, in this work, to endogenize and interrelate governance and culture at the "micro" societal scale: the opportunity provided by large collections of online communities to track and compare thousands of similar but largely independent small-scale sovereign social systems.

Figure 1

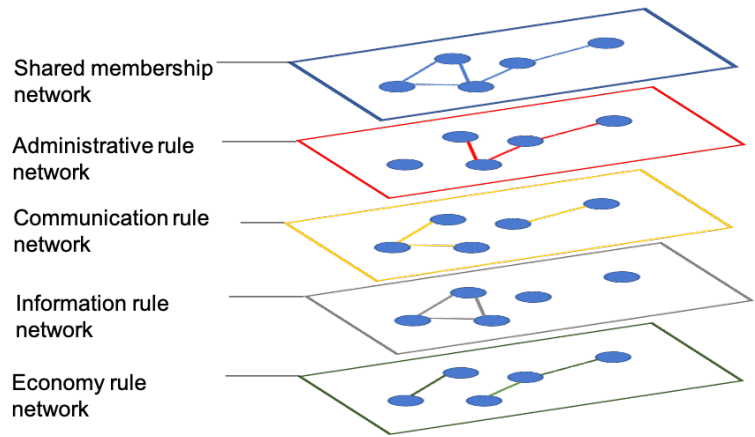

**Figure 1 A multiplex network representation of many servers' patterns of shared membership and common institutional structure.** We layered the community network and four rule networks on top of each other to construct a multiplex network. The nodes in each layer, which signify the servers, are the same, whereas the relations in different network layers have different meanings. We consider servers to be culturally similar if they attract the same type of user. If the deletion or formation of links in, for example, the shared membership network influences the deletion or formation of links in a rule network, above baseline, then there is evidence that culture affects institutions, rather than the other way around. The position of the networks in the figure does not indicate the order of layers in the multiplex. Relational effects might occur between any of the two layers in the multiplex, but we focus on only four comparisons, those of the shared membership network with each type of rule network.



Figure 2

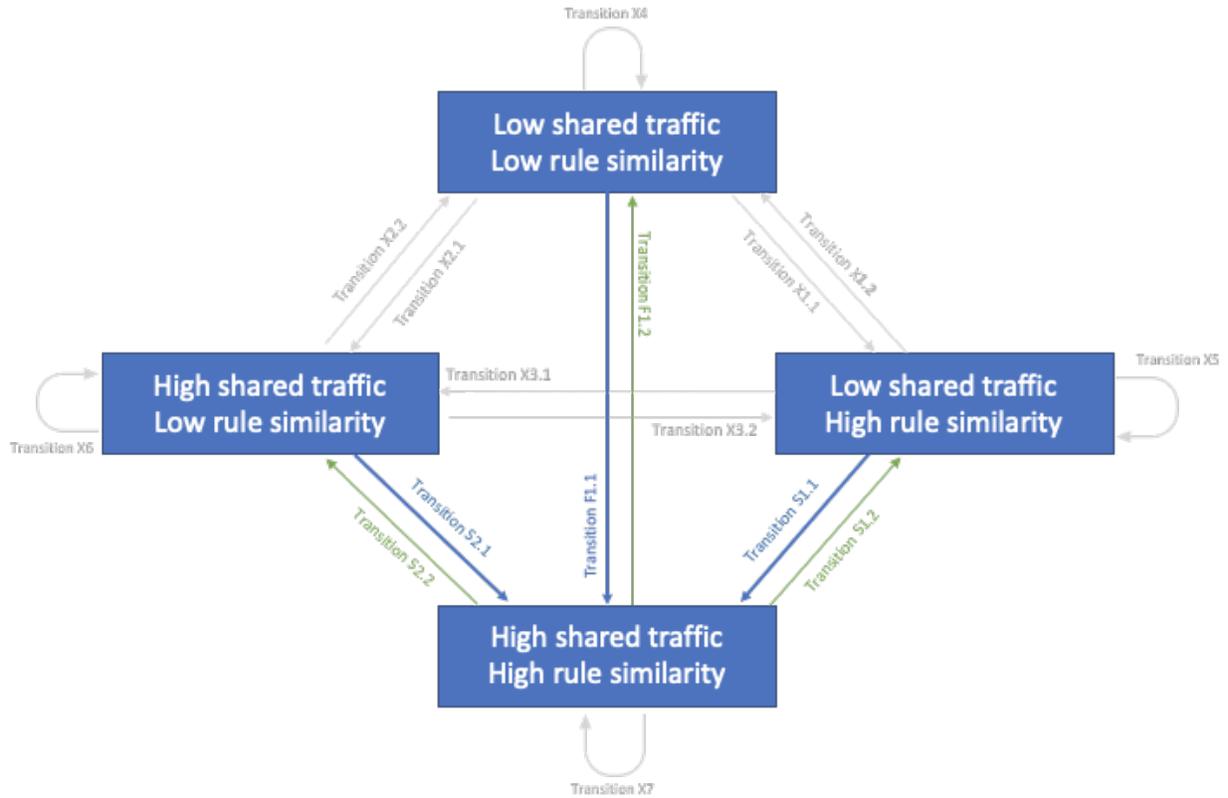

**Figure 2 Dynamics of culture/institution interactions in terms of patterns of changes in the types of link that can connect two nodes** We illustrate all possible transitions of interactions between shared traffic and rule similarities throughout the process of community development, which renders visible institutional processes over several timescales, and pits predictions of institutional effects on culture and cultural influence on institution. F represents fast-timescale transitions (in which the extent of shared governance and membership changed simultaneously in a month); S represents slow-timescale transitions (in which one changed before the other); X represents transitions that are irrelevant to the hypotheses.



Figure 3

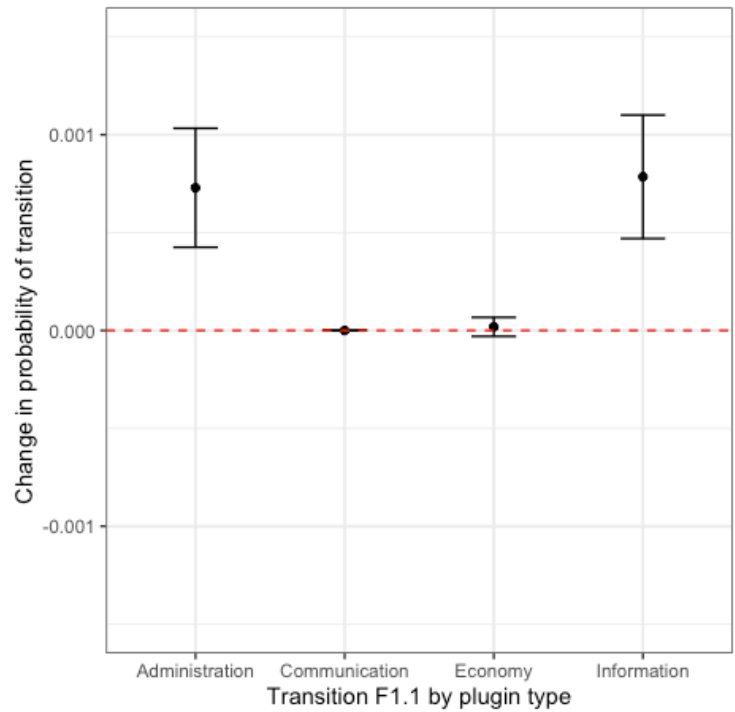

**Figure 3 Evidence for correlations between similarities in cultural and institutional development**  Between two communities, a common trend was for an increase from communities with low similarity to high similarity in both membership and rule type, in the administrative and informational rules among the four rule types (Transition F1.1). For these rule types, does similarity in shared traffic drive increased similarity in rule type or does similarity in rule type drive increased similarity in shared traffic?  This figure represents those link changes from Figure 1 that demonstrate the existence of "cross-layer" influence between our rule and cultural networks, but do not indicate a direction for that influence.



Figure 4

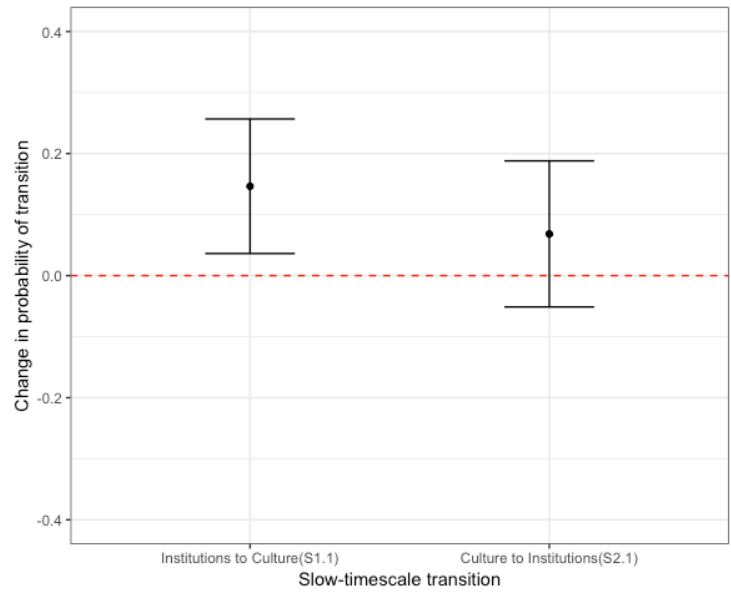

**Figure 4 Institutional effects over cultural effects comparing to the other way round in the administrative rule type** Among plugins that focus on administrative rule type, two communities with the same focus on administrative capacity are more likely to have shared traffic, while two communities with different degrees of reliance on this rule type are less likely to have a shared culture.



Figure 5

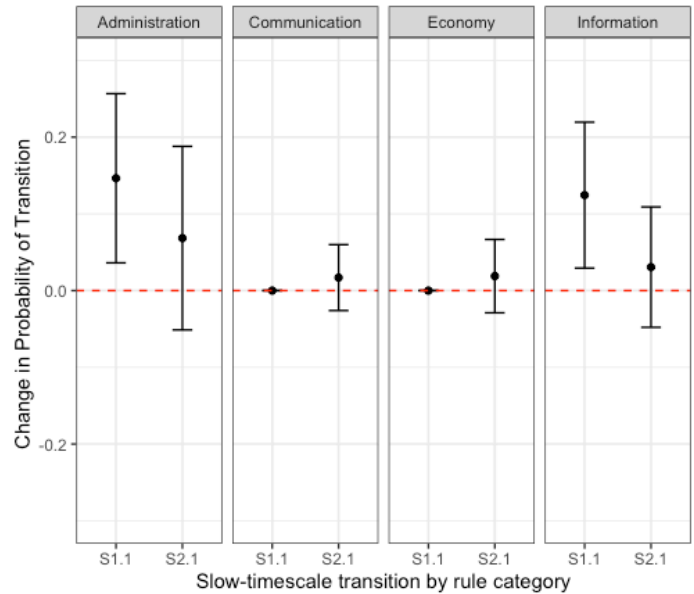

**Figure 5 Institutional effects over cultural effects in the administrative and informational rule category.** We test the effects between institution and culture for all four rule types, finding influence from institutional to cultural similarity (The relationship shown in Figure 4 is repeated in the first panel to provide an aid to comparison). We produce the same effect for administrative and informational rules, but find no effect for communication or economic rules.



Supporting Figure 1

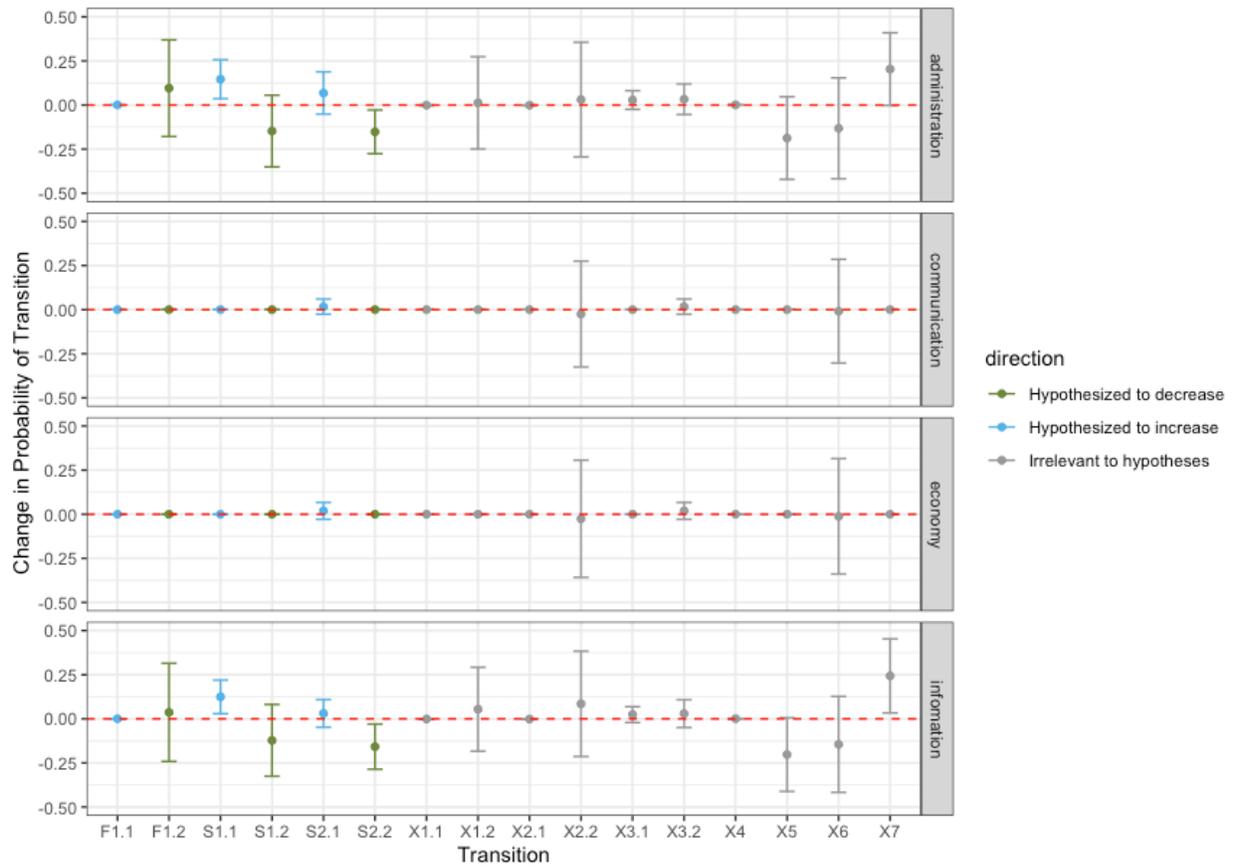

**Supporting Figure 1. All transition spillover probability within four rule categories.** Across governance capabilities, we produce the effects of all transitions group by relevance to the hypotheses.



Supporting Figure 2

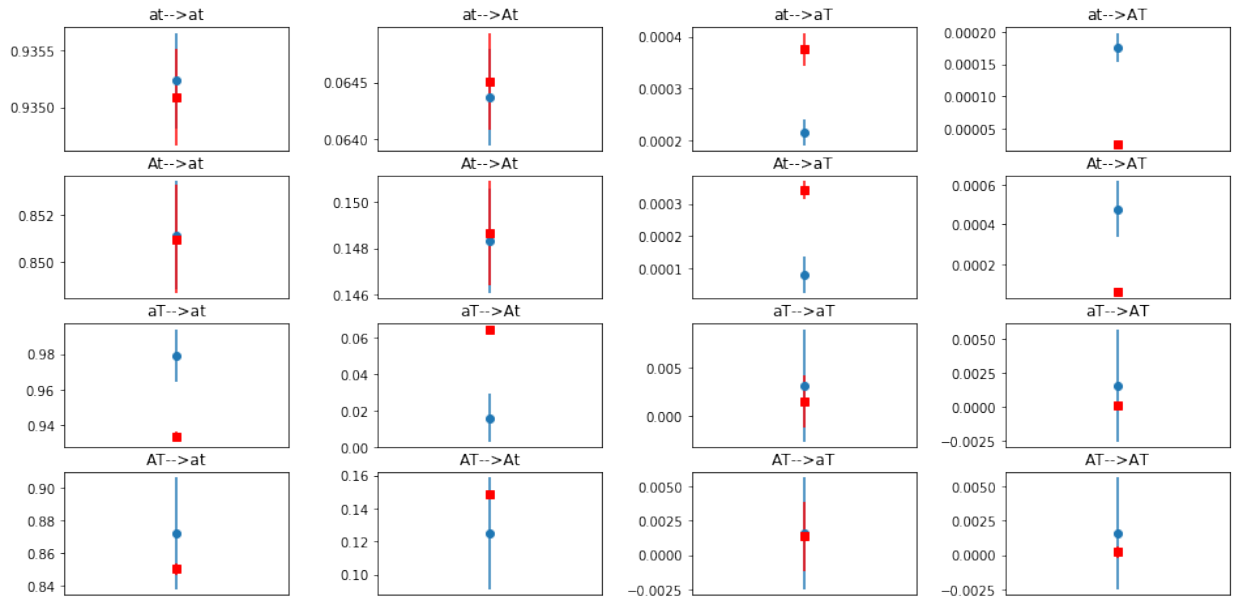

***Supporting Figure 2. Comparing mutual effects of the administrative rule network and the community network, compared to null model statistics*** This set of figures shows the state transition probabilities estimated for a multiplex Markov chain (blue error bar) and those of the null Markov model (red error bar), where states are the presence or absence of links across different layers of the network.  This figure presents the same result as the first row of Supporting Figure 1, the difference being that this figure shows both the null and multiplex transition probabilities, rather than their difference. We label the absence of a link with a lower-case letter. In the panel titles, *lower-case a* represents *absent* ties in administrative rule network, and *lower-case t* represents *absent* ties in community traffic network, while we indicate the *presence* of a link with an *upper-case* letter, so that *A* represents existing ties in administrative rule network and *T* existing ties in community traffic network. For example, an *at* → *AT* transition is a transition from no ties in both networks to existing ties in both. We observe a positive (negative) effect for a transition when the confidence interval around the multiplex Markov chain's estimated transition probability (blue error bar) lies entirely above (below) the confidence interval of the null (non-multiplex) Markov model's transition probability (red error bar).



Supporting Figure 3

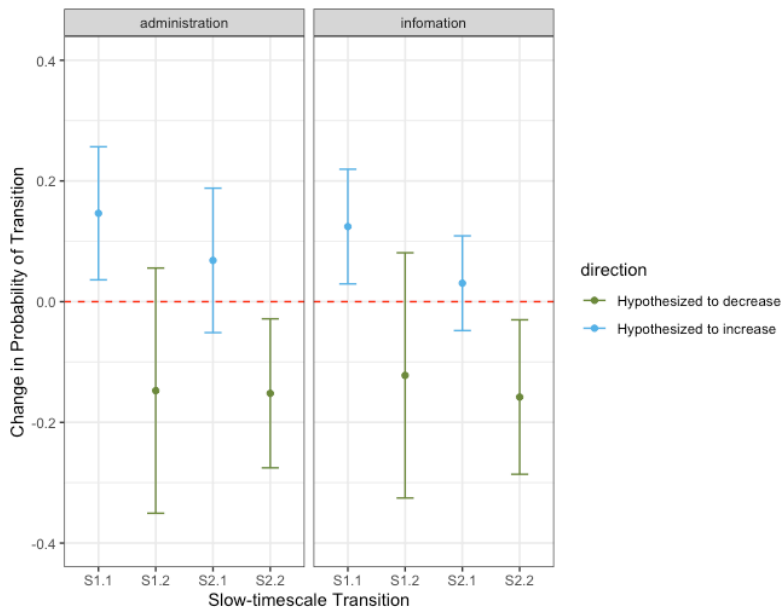

***Supporting Figure 3. For administrative and information capacities, we also produced negative effects for the opposite-direction transitions.*** The transition S1.2 and S 2.2 offer complementary result to S1.1 and S2.1 The positive effects in S1.1 and S2.1 indicate that servers similar in one dimension are more likely to become similar in the other dimension. Correspondingly, servers that are similar in one dimension are less likely to stop being similar in another dimension.